\documentclass[useAMS]{mn2e_modmargin}
\usepackage{amsmath}
\usepackage{url}
\usepackage{amsfonts}
\usepackage{amsbsy}
\usepackage[dvips]{graphics}
\usepackage{subfigure}
\usepackage{verbatim}
\usepackage{amssymb}
\usepackage{amsbsy}

\renewcommand{\d}{\ensuremath{\partial}}

\newcommand{\ey}{\ensuremath{\mathbf{e}_{y}}}
\newcommand{\ez}{\ensuremath{\mathbf{e}_{z}}}

\newcommand{\ii}{\ensuremath{\text{i}}}

\newcommand{\ee}{\text{e}}
\newcommand{\iii}{\text{i}}

\title[The ballistic transport instability I]{The ballistic transport
instability in Saturn's rings I:
formalism and linear theory}
\author[Latter, Ogilvie \& Chupeau]{Henrik N. Latter$^{1}$\thanks{E-mail:
    hl278@cam.ac.uk},
   Gordon I. Ogilvie$^{1}$\thanks{E-mail:
    gio10@cam.ac.uk},
   Marie Chupeau$^{1,2}$\thanks{E-mail:marie.chupeau@ens-cachan.fr} \\
$^{1}$ DAMTP, University of Cambridge, CMS, Wilberforce Road,
Cambridge CB3 0WA, UK\\
$^{2}$ ENS Cachan, 61 Avenue du Pr\'esident Wilson, 94230 Cachan, France}
\date{}

\begin{document}

\maketitle

\begin{abstract}
   Planetary rings sustain a continual bombardment
    of hypervelocity meteoroids that erode the surfaces of ring
    particles on time scales of $10^5$--$10^7$ years.
    The debris ejected from such impacts
    re-accretes on to the ring, though
    often at a slightly different orbital radius from the point of
    emission.  This `ballistic transport' leads to a rearrangement of
    the disk's mass and angular momentum, and gives rise
    to a linear instability that generates structure on
    relatively large scales. It is likely that the 100-km undulations
    in Saturn's inner B-ring and the plateaus and 1000-km waves in
    Saturn's C-ring are connected to the nonlinear saturation of the
    instability. In this paper the physical problem is reformulated so
    as to apply to a local patch of disk (the shearing sheet).
    This new streamlined model helps facilitate our
    physical understanding of the instability, and also makes more
    tractable the analysis of its nonlinear dynamics.  We concentrate
    on the linear theory in this paper, showing that the instability
    is restricted to a preferred range of intermediate wavenumbers and
    optical depths. We subsequently apply these general results to the
    inner B-ring and the C-ring and find that in both regions the
    ballistic transport instability should be near marginality, a fact
    that may have important consequences for its prevalence and
    nonlinear development. Owing to damping via self-gravity wakes,
    the instability should not be present in the A-ring.  A following
    paper will explore the instability's nonlinear saturation and how
    it connects to the observed large-scale structure.

\end{abstract}


\section{Introduction}

Like all solar-system bodies, the component particles of planetary
rings must endure a hail of interplanetary meteoroids impacting at
speeds on the order of $10$~km~s$^{-1}$. Collectively the most erosive
projectiles lie in the size range of $10^{-2}$--$10^{-1}$~cm, and it
is estimated that Saturn sweeps up such particles at a rate $\gtrsim$
10~kg~s$^{-1}$ (Durisen 1984; Ip 1984; Cuzzi \& Durisen 1990, hereafter
CD90).  Hypervelocity impacts liberate
significant amounts of material from ring particles (some
$10^3$--$10^5$ times the mass of each impactor), and these ejecta
re-accrete on to the ring, typically at a different radial location
from where they started. The exchange of ejecta between nearby
regions, referred to as `ballistic transport', facilitates a
redistribution of mass and angular momentum on length-scales
$l_\text{th}\sim 10-10^3$~km and times $t_\mathrm{e}\sim
10^5-10^7$ yrs that should control the large-scale evolution of
Saturn's rings.  Indeed, these scalings suggest that, over the age of
the solar system,
$10^2$--$10^4$ times the mass of the
current rings has been transported in this way.

Previous theoretical work shows that the ballistic transport
process is adept at reshaping pre-existing gradients in surface
density and composition. It can sharpen edges, such as those at the
inner boundaries of Saturn's A and B-rings, generate the ramp
features at the feet of those edges, and influence global colour
gradients (Ip 1983; Lissauer 1984; Durisen 1984; Durisen et al.~1989,
hereafter D89; Durisen et al.~1992; Cuzzi \& Estrada 1998; Charnoz et
al.~2009). But ballistic transport can also produce structure
spontaneously from a homogeneous ring via a linear instability. This
`ballistic transport instability'
is thought to drive the 100-km wavetrains in the inner
B-ring, and possibly the 100-km-wide plateaus and the low-amplitude
1000-km undulations in the C-ring (Durisen et al.~1992; Durisen 1995,
hereafter D95; Colwell et al.~2009; Charnoz et al.~2009).  It is to
the basic theory of ballistic transport and the linear instability that this paper will be
devoted.

The dynamics of ballistic transport has been successfully described by
a detailed global model constructed by Durisen and coworkers in the
1980s and 1990s (D89; CD90; Durisen et al.~1992; D95; Durisen et
al.~1996; Cuzzi and Estrada 1998). The model
incorporates a great many physical processes, but the mathematical
formalism can be unwieldy and thus potentially obscure the fundamental
physics. 
In this paper we omit extraneous details and construct a simpler
model that is easier to work with, yet remains sufficiently accurate.
It is, in fact, almost identical to the leading-order Durisen
formalism when expanded in the small parameter $\varrho=
v_\mathrm{e}/v_\mathrm{c} \ll 1$, where $v_\mathrm{e}$ is the typical
relative speed of the ejecta, and $v_\mathrm{c}$ is the orbital speed
of the ring particles. (Note that Durisen and coworkers denote this
ratio by $x$, which we reserve for a radial coordinate.)
As a result, our model is local (the
shearing sheet), and this makes the ejecta orbital dynamics
easy to describe. The local formalism also permits the resulting transport terms
of mass and angular momentum to be manipulated into one-dimensional
integrals in convolution form. Being especially amenable to Fourier
analysis, the linear theory is transparent and nonlinear simulations are comparably
straightforward.
Overall, the simpler formalism facilitates our physical
understanding of the instability, and permits us to bring to bear
the techniques of nonlinear dynamical systems.
Additional physics can always be added later to sharpen
the quantitative comparison with observations.

In this paper we concentrate on the linear analysis of the system.
It yields a simple instability criterion, which reveals instability
is facilitated, in particular, by the decrease in the ejecta emission rate near optical
depths of 0.5. This drop corresponds to a transitional regime in which
the disk becomes sufficiently dense that some of the liberated ejecta
are reabsorbed by neighbouring particles rather than sent into orbit.
In agreement with D95, we find that instability is suppressed at very
low and high optical depths. But we also see that it is suppressed at
very long and short wavelengths, and that unstable modes can propagate
either radially inwards or outwards, depending on wavenumber.  Our
analysis frames the problem in terms of two dimensionless parameters:
the mean optical depth $\tau_0$ and the `ballistic
Prandtl number' $\mu$, which describes the relative efficiency of
  mass redistribution caused by viscous stresses versus that caused by
  ballistic transport. Most of the uncertainties in the problem are
 packaged into $\mu$, which nevertheless can be tightly
constrained.
In the A-ring $\mu$ is
relatively large, because of strong self-gravity wakes, and therefore viscous
diffusion smears out potentially unstable modes.  In both the inner
B-ring and the C-ring $\mu$ is smaller and linear modes may grow.
However, in these two cases instability is near marginality, within the
uncertainties, because $\tau_0$ is small or large, respectively.
Though instability is still likely to occur, the fact that the system
is near marginality will influence its nonlinear development,
potentially leading to low-amplitude saturation or bistability. A
weakly nonlinear analysis of these cases, together with fully
  nonlinear numerical simulations, will be presented in following
work.

The paper is organised as follows. In Section~2 we detail the
mathematical formalism that we use to describe the ballistic transport
process, culminating in the governing equation for optical depth,
Eq.~\eqref{taueq}. There we also discuss the functional forms for
the rate of emission, the probability of absorption and the ejecta
throw distribution, which we draw from fits to the numerical
calculations of CD90. The linear stability analysis follows in
Section~3, in which we present growth rates and a general instability
criterion with application to Saturn's A, B, and C-rings.
We discuss these results and conclude in Section~4.

\section{Mathematical formalism}

We consider a local model of a particulate ring, the shearing sheet (Fig.~1),
which is a convenient representation of relatively small-scale
dynamics in a differentially rotating disk (Goldreich \& Lynden-Bell
1965).  Instead of dealing with the entirety of a ring system, with
its unconstrained global structure and boundary conditions, we
concentrate on a small patch of disk centred on a fiducial radius
$r_0$ and moving on a circular orbit with angular velocity
$\Omega_0=\Omega(r_0)$.  The small patch can then be described by a
Cartesian coordinate system $(x,y,z)$, with $x$ and $y$ pointing in
the radial and azimuthal directions respectively, and its differential
rotation represented by a combination of uniform rotation,
$\Omega_0\,\ez$, and linear shear, $\mathbf{v}=-S_0\,x\,\ey$, where
$x=r-r_0$. The rate of orbital shear is $S_0= -r_0(d\Omega/dr)_0$ and
is equal to $\tfrac{3}{2}\Omega_0$ in a Keplerian disk. The
shearing-sheet approximation introduces fractional errors of order
$\lambda/r_0$, where $\lambda$ is the characteristic length-scale of
the dynamics we want to describe.  Therefore, its descriptions of
100-km-long waves in the B-ring and plateaus in the C-ring yield an error $\sim
10^{-3}$, whereas the very slow 1000-km undulations in the C-ring give
$10^{-2}$. Similarly, the ballistic transport process can be described
adequately by the shearing sheet, as the errors introduced scale as
$l_\text{th}/r_0\sim 10^{-3}$--$10^{-2}$.

\begin{figure}
\begin{center}
\scalebox{0.6}{\includegraphics{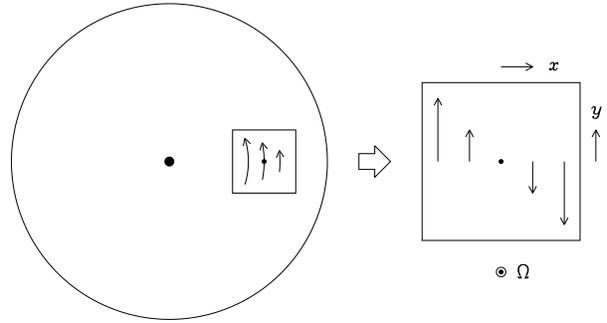}}
\caption{Diagram showing the rationale and main features of the
  shearing sheet model. A small patch of disk is isolated and treated
  as a Cartesian sheet subject to rotation and a shear flow. Terms
  arising from the cylindrical geometry are dropped and the
  coordinates $x$ and $y$ point in the radial and azimuthal direction
  respectively.}
\end{center}
\end{figure}

Let $\sigma(x,t)$ be the surface mass density of the ring, assumed
from the outset to be axisymmetric. We take the particle size
distribution to be fixed and assume there exists a unique relation
between $\sigma$ and the normal optical thickness $\tau$.  Mass
conservation furnishes us with the following evolution equation for
the ring:
\begin{equation}\label{SigmaGeneral}
\d_t \sigma  + \d_x (\sigma\,u_x) = \mathcal{I} - \mathcal{J},
\end{equation}
where $\mathcal{I}(x,t)$ is the rate at which mass is gained, per unit
area, via ballistic transport from other radii, and $\mathcal{J}(x,t)$
is the rate at which mass is lost. The radial drift speed within the
ring is denoted by $u_x$ and is instigated by both viscous stresses
and the ballistic transport of angular momentum. In the following
subsections we derive a formalism that supplies us with convenient
expressions for $\mathcal{I}$, $\mathcal{J}$ and the local
mass flux $\sigma\,u_x$.

\subsection{Characterising the properties of ejecta emission and absorption}

\subsubsection{Rate of emission, $R$}

The local rate at which mass is liberated from the ring per unit area
as a result of meteoritic bombardment is the erosion rate
$R[\sigma(x,t)]$. This suggests a local gross erosion time
$t_\mathrm{e}= \sigma/R[\sigma]$, which corresponds to the time it
would take for the ring to be completely destroyed, in the absence of
ejecta recycling. At $\tau=1$, an estimate for $t_\mathrm{e}$ is
\begin{equation} \label{te}
t_\mathrm{e} = 10^6 \left(\frac{10^4}{Y}\right)\left(\frac{\sigma}{100\,\text{g
cm}^{-2}}\right)\, \text{yr},
\end{equation}
where the yield $Y$ is the ratio of liberated mass in a meteoroid
impact to the mass of the impactor (D95). The value of $Y$ depends
sensitively on the strength and composition of the ring particle's
surface, with `softer' particles releasing more material and thus
taking larger $Y$. Though the physical state of ring particles is
poorly constrained, terrestrial laboratory experiments give $Y\approx
10^3$--$10^5$ for hard water-ice/silicate targets struck by impactors
travelling at speeds $\sim 0.1$--$10$~km~s$^{-1}$ (Lange \& Ahrens
1987; Frisch 1992; Koschny \& Gr\"un 2001a). 
Note that these estimates pertain only to cratering (non-disruptive) impacts, which are the most common outcome in Saturn's rings. 
Disruptive impacts produce higher effective yields, 
the cumulative effects of which may influence regions of the rings in which particles are smaller and hence more prone to destruction (Estrada and Durisen 2010).

\subsubsection{Throw distribution, $f$}

Liberated material travels in a slightly inclined and slightly
eccentric Keplerian orbit that intersects the ring plane at the
location of emission, $x$, and a distant location, $x+\xi$. The radial
distance $\xi$ between these two locations depends on the ejection
velocity, the statistics of which is summarised in a `throw
distribution' $f(\xi)$, normalised such that $\int f(\xi)\,
\mathrm{d}\xi=1$. (Unless otherwise specified, all integrals are
carried out from $-\infty$ to $\infty$.) Thus $f(\xi)\,\mathrm{d}\xi$
is the proportion of ejecta that travels a radial distance between
$\xi$ and $\xi+\mathrm{d}\xi$ during its orbit. The distribution
should exhibit a characteristic length-scale, the `throw length'
$l_\text{th}$. By considering the dynamics of ejecta trajectories
(Subsection 2.2.1; Appendix~A), we set
\begin{equation} \label{lth}
l_\text{th}=
4\,r_0\,\varrho = 2\times 10^2\,\left(\frac{v_\mathrm{e}}{10\,\text{m
s}^{-1}}\right)\left(\frac{r_0}{10^5\,\text{km}}\right)^{3/2}\,\text{km},
\end{equation}
where $\varrho= v_\mathrm{e}/(r_0\Omega)$ and here $v_\text{e}$ refers
to the mean ejection speed. Thus $l_\text{th}$ is the maximum throw
possible for the given mean ejection speed $v_\text{e}$.
The magnitude of $v_\text{e}$,
like the yield $Y$, depends closely on the physical state of the ring
particle's surface, with `softer' particles possibly taking $1$ m~s$^{-1}$
and `harder' particles taking $100$ m~s$^{-1}$ (D89; Frisch 1992; Koschny \&
Gr\"un 2001b).
Consequently, the small parameter $\varrho$ varies between $10^{-4}$
and $10^{-2}$, while $l_\text{th}$ lies between 10
and 1000~km, which encompasses the scales of the phenomena we aim to
describe.

\subsubsection{Probability of absorption, $P$}

During their orbit, ejecta may be reabsorbed by the ring at either the
radius of emission $x$ or at the distant intersection radius $x+\xi$.
The probability that an ejectum is absorbed at $x+\xi$ we denote by
$P[\sigma(x+\xi,t),\sigma(x,t)]$. The reason for allowing $P$ to
depend on the surface density at the emitting radius $x$, as well as
that at the distant radius $x+\xi$, is to account for extended
excursions whereby the ejectum passes through the ring plane multiple
times. This outcome is likely only in optically thin regions of the
disk, and thus the second dependence may be dropped for optically
thicker regions. Note that the ballistic transport of individual ejecta occurs on an orbital
period $\sim 10$~hr, much less than the erosion time $t_\mathrm{e}$.
Hence the time required to execute multiple orbits can be neglected.

\subsection{Direct mass transport}

Now that we have introduced $R$, $f$ and $P$ we can construct the mass
gain and loss terms $\mathcal{I}$ and $\mathcal{J}$. The rate at which
mass is lost from a radius $x$ is $R[\sigma(x,t)]$, while the
proportion of this mass that travels to an annulus a distance $\xi$
away and of thickness
$\mathrm{d}\xi$ is $f(\xi)\,\mathrm{d}\xi$. Finally, the fraction of
this mass that is ultimately absorbed by this annulus is
$P[\sigma(x+\xi,t),\sigma(x,t)]$. Now if we sum over all such annuli
we obtain the total loss rate
\begin{equation}\label{J1}
\mathcal{J}(x,t) = R[\sigma(x,t)]\,\int
P[\sigma(x+\xi,t),\sigma(x,t)]\,f(\xi)\,\mathrm{d}\xi.
\end{equation}
A similar argument accounts for the total gain of mass at $x$ from all
neighbouring annuli:
\begin{equation} \label{I1}
\mathcal{I}(x,t) = \int
R[\sigma(x-\xi,t)]\,P[\sigma(x,t),\sigma(x-\xi,t)]\,f(\xi)\,\mathrm{d}\xi.
\end{equation}

The gain and loss integrals, though potentially complicated, benefit
from being
one-dimensional. The integrals in the formalism of D89, on the other
hand, are three dimensional and cover the two emission angles of the
ejecta as well as the speed of emission. These may be more accurate but
are awkward in analytic work, and costly in numerical
simulations. Moreover, the three degrees of freedom in D89 should map
approximately to a single throw distance $\xi$ in the case of
$\varrho\ll 1$. In
Appendix~A we show in detail how the two formalisms join up.

\subsection{Angular momentum transport}

Both the ring's viscous stresses and the radial redistribution of ejecta
lead to angular momentum transport and a consequent radial drift of
material. In order to calculate the mass flux associated
with the slow radial motion, we need to
sketch out the trajectories of the ejecta in the shearing sheet.

\subsubsection{Ejecta trajectories in the shearing sheet}

\begin{figure*}
\begin{center}
\scalebox{0.5}{\includegraphics{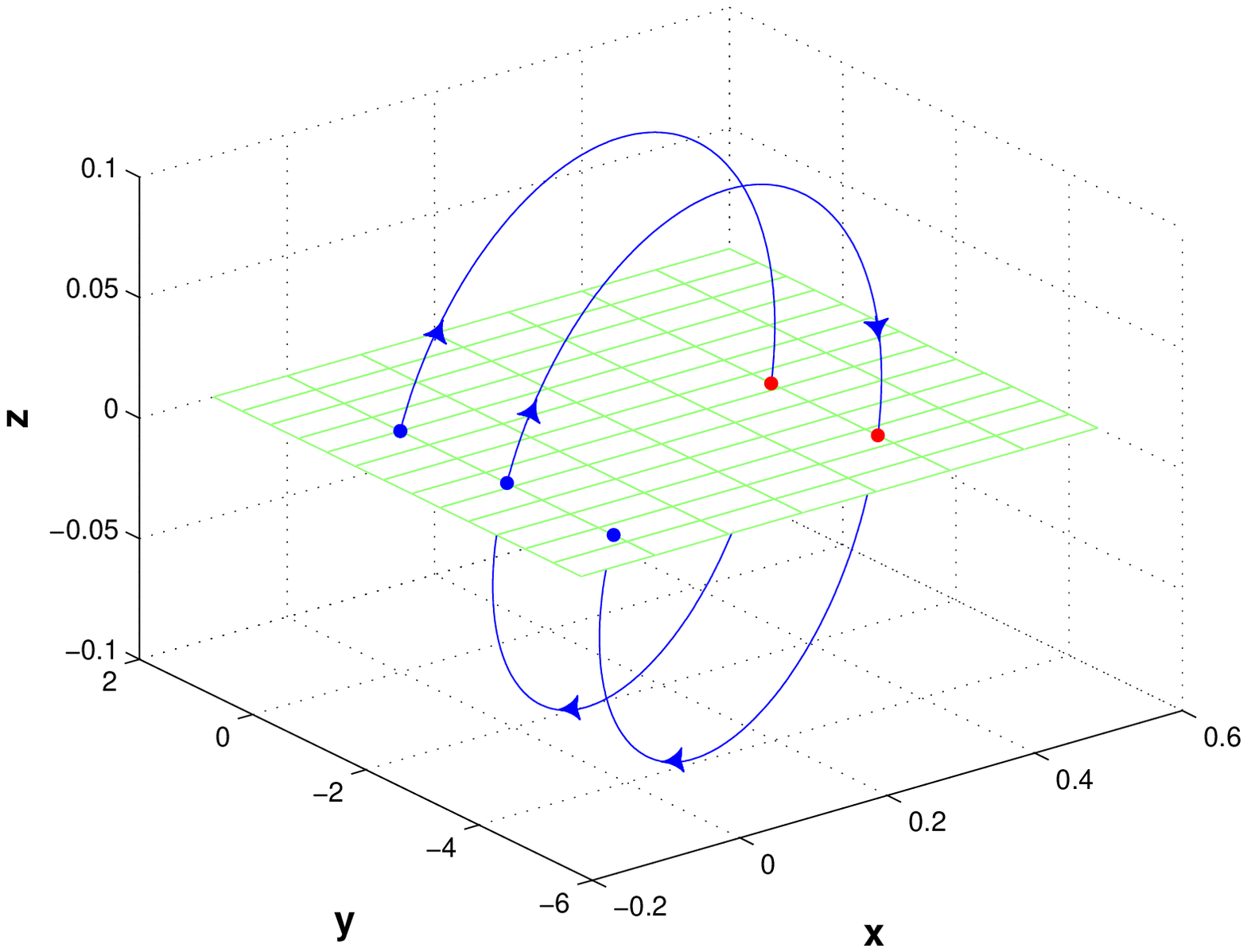}}
\scalebox{0.4}{\includegraphics{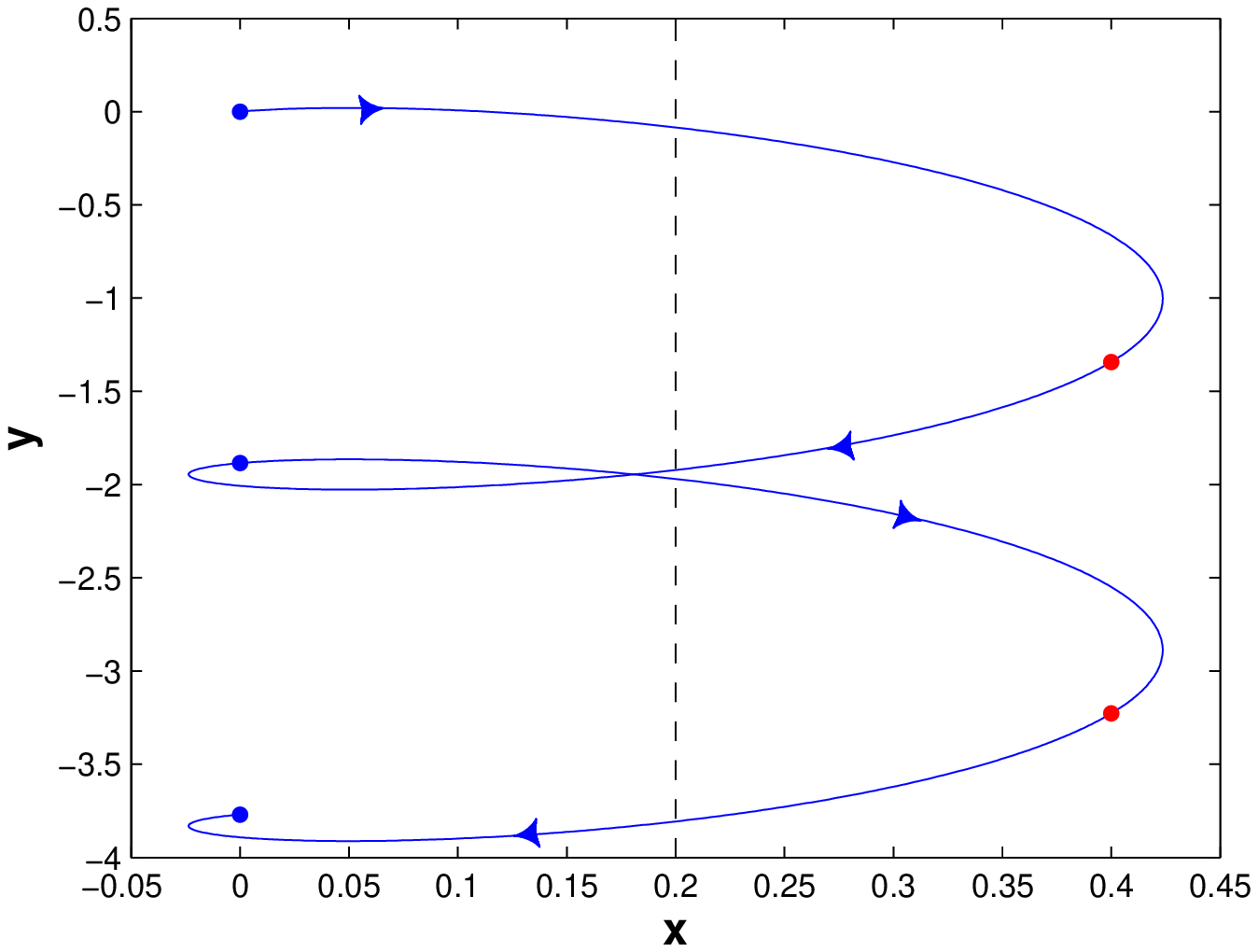}}
 \caption{A representative ejectum trajectory begun at $(0,0,0)$ and
   evolved forward for 4 ring-plane crossings according to Eqs
   \eqref{xtraj}--\eqref{ztraj}.
    The initial ejection
   speed is $(u,v,w)=(0.1,0.1,0.1)$. The first panel describes the
 trajectory in three dimensions with the green grid indicating the
 ring plane. The second panel projects this trajectory on to the
 $(x,y)$ plane. The dashed black line denotes the radius that
 corresponds to the ejectum's angular momentum. In both figures the
 axes have been stretched relative to each other in order to
 bring out more clearly the details of the motion.}\label{Fig:traject}
\end{center}
\end{figure*}

The orbit of a representative ejectum $[x(t),y(t),z(t)]$ in the shearing sheet
obeys
the Hill equations
\begin{align}
&\ddot x - 2\Omega_0 \dot y - 3 \Omega_0^2 x = 0, \\
&\ddot y + 2 \Omega_0 \dot x = 0, \\
&\ddot z + \Omega_0^2 z = 0,
\end{align}
where an overdot indicates a time derivative.  (Throughout this paper we
neglect non-Keplerian effects arising from planetary oblateness.) If the
ejectum is
thrown from the ring at $t=0$ from position $(0,0,0)$ (without loss of
generality) and with
ejection velocity $(u,v,w)$, it will undergo the following trajectory:
\begin{align}
&x= \frac{1}{\Omega_0}\left[ u\, \sin \Omega_0 t + 2v\,(1-\cos\Omega_0
  t)\right], \label{xtraj}\\
&y= \frac{1}{\Omega_0}\left[-2u(1-\cos\Omega_0 t) + v(4\sin\Omega_0 t - 3
  \Omega_0 t)   \right], \\
&z = \frac{w}{\Omega_0}\sin \Omega_0 t. \label{ztraj}
\end{align}
The ejectum will first return to the ring plane half an orbit later, at
$t=\pi/\Omega_0$, at
position $(4v,-4u-3\pi v, 0)/\Omega_0$ and with velocity $(-u,-7v,-w)$. If it
is not reabsorbed at this point it will continue its orbit, meeting the
ring plane again at $t=2\pi/\Omega_0$ at position $(0,-6\pi v,0)/\Omega_0$ with
its original
velocity, $(u,v,w)$. A representative trajectory is described in
Fig.~\ref{Fig:traject} in three dimensions and also projected in the
$(x,y)$ plane. The blue dots denote ring crossings at even integer
multiples of $t=\pi/\Omega$, and red dots the ring crossings at odd
multiples.

In the shearing sheet context, the role of (specific) angular momentum is
played
by the (specific) canonical azimuthal momentum $p_y= \dot y + 2\Omega_0 x$, a
quantity that
is conserved in the Hill equations, and which differs from the
$y$-velocity because of
 the Coriolis force. For a particle in a circular orbital motion
at a fixed $x$ we have $\dot y = -\tfrac{3}{2}\Omega_0\,x$ and thus
$p_y=\tfrac{1}{2}\Omega_0 x$.  The uniform radial gradient of this quantity
in the shearing sheet corresponds
(apart from a factor of $r_0$) to the local radial gradient of
specific angular
 momentum among the family of circular Keplerian orbits. For the ejectum orbit
considered above, $p_y=v$,
which can be seen from considering the conditions at $t=0$ at the
moment
the particle is
launched. But this value corresponds to a circular orbit located at
$x=2v/\Omega_0$, exactly halfway between its two ring-crossing
radii. The radius of this orbit is described by the dashed black line
in Fig.~\ref{Fig:traject}b.
 So when an ejectum
 is launched, it acquires a small amount (positive or negative) of
 angular momentum; subsequently, it oscillates about the radius
 associated with its new angular momentum and is absorbed at one or other
radial extremum. Translating this result
 into the terminology of the previous subsection, we can then say that
 material emitted from $x$ and with throw distance $\xi$ has
 $p_y=\tfrac{1}{2}\Omega_0\,(x+\xi/2)$.

\subsubsection{Radial mass drifts due to ballistic transport and viscosity}

We now consider the evolution of the angular momentum analogue of the
ring itself, $\sigma\,p_y$. The conservation law for
this quantity in an axisymmetric ring is
\begin{equation} \label{sigmapy}
\d_t \left[\sigma(x,t)p_y\right] + \d_x Q = \overline{\mathcal{K}} -
\overline{\mathcal{L}},
\end{equation}
where $\overline{\mathcal{K}}$ and $\overline{\mathcal{L}}$ are the
rate of gain and loss of
 angular momentum respectively at $x$ due to ballistic transport, and
 $Q$ is the local momentum flux density. It
 is equal to
\begin{equation}
Q = \sigma\,u_x\,p_y - \Pi_{xy},
\end{equation}
where $\Pi_{xy}$ is the $xy$ component of the viscous stress tensor.
 As ring material follows circular orbits
predominantly, we take $p_y= \tfrac{1}{2}\Omega_0 x$ from now.

The two ballistic transport terms $\overline{\mathcal{K}}$ and $\overline{\mathcal{L}}$ are straightforward to construct.
Consider ejecta released at location $x$ and absorbed at $x+\xi$ in an annulus
of thickness $\text{d}\xi$. The (specific) angular momentum of such
ejecta is  $\tfrac{1}{2}\Omega_0 (x+\xi/2)$. As before, the rate of its
emission is $R[\sigma(x,t)]$, and the proportion that travels to the
annulus is  $f(\xi)\,\mathrm{d}\xi$. Finally, the fraction that is
absorbed at $x+\xi$ is $P[\sigma(x+\xi,t),\sigma(x,t)]$. Summing over
all the neighbouring annuli yields the rate of angular momentum loss at
$x$:
\begin{align}
\overline{\mathcal{L}}(x,t) &= R[\sigma(x,t)]\int
P[\sigma(x+\xi,t),\sigma(x,t)]\,f(\xi)\, \notag \\
 & \hskip3cm \times\tfrac{1}{2}\Omega_0
(x+\xi/2)\,\mathrm{d}\xi.
\end{align}
The gain rate can be constructed in a similar way:
\begin{align}
\overline{\mathcal{K}}(x,t)&= \int
R[\sigma(x-\xi,t)]\,P[\sigma(x,t),\sigma(x-\xi,t)]\,f(\xi) \notag \\
 & \hskip3cm \times\tfrac{1}{2}\Omega_0\,(x-\xi/2)\,\mathrm{d}\xi. \label{totK}
\end{align}

The nature of the viscous stress $\Pi_{xy}$ in cold and dense particulate rings,
such as Saturn's, is nontrivial and comprises various components,
each of which can deviate from the familiar Newtonian prescription
(see e.g.\ Latter \& Ogilvie 2006, 2008; Schmidt et al.~2009). In the
inner rings viscous transport is dominated by the `collisional stress'
(Shukhman 1984; Araki \& Tremaine 1986;
Wisdom \& Tremaine 1988), whereas stresses arising from self-gravity
wakes prevail in the A-ring
(Salo 1992; Daisaka et al.~2001, Yasui et al.~2012). The internal processes governing both
operate on the orbital time-scale, which is much shorter
than $t_\mathrm{e}$. It hence makes sense to treat the stress
in the diffusion approximation and to introduce an effective viscosity $\nu$.
The
viscous mass flux then becomes
\begin{equation}
\Pi_{xy} = -\frac{3}{2}\Omega_0\nu\sigma.
\end{equation}
Generally, the viscosity $\nu$ may be considered a function of $\sigma$ or
optical thickness $\tau$. For simplicity, we treat $\nu$ here as a constant, although the linear theory presented in this
paper is easily adapted, as we describe below, to the more realistic situation of a density-dependent viscosity.

Finally, by subtracting from Eq.~(\ref{sigmapy}) $\tfrac{1}{2}\Omega_0 x$ times
the mass-conservation equation
\eqref{SigmaGeneral}, we can obtain an expression for the mass flux density
$\sigma\,u_x$ that appears in the governing equation \eqref{SigmaGeneral},
\begin{equation}
\sigma\,u_x = -\tfrac{1}{2}(\mathcal{K}+\mathcal{L}) -3\d_x (\nu\,\sigma).
\end{equation}
This expression introduces new transport integrals defined
through
\begin{align}
\mathcal{K} &= -\frac{4}{\Omega_0}(\overline{\mathcal{K}} -
\tfrac{1}{2}\Omega_0x\, \mathcal{I})
\end{align}
and
\begin{align}
\mathcal{L} &= \frac{4}{\Omega_0}\left(\overline{\mathcal{L}} -
\tfrac{1}{2}\Omega_0x\, \mathcal{J}\right).
\end{align}
Thus $\mathcal{K}$ and $\mathcal{L}$ are identical to $\mathcal{I}$ and
$\mathcal{J}$ but with an extra factor of $\xi$ in their integrands.

Our governing equation for the surface density is then
\begin{equation}\label{dsigmadt}
\d_t \sigma  = \mathcal{I} - \mathcal{J} + \tfrac{1}{2}
\d_x(\mathcal{K}+\mathcal{L}) + 3\nu\d_x^2\sigma.
\end{equation}
As is well known, the viscous transport of angular momentum leads to a
radial spreading of mass.
The mass diffusion coefficient in a Keplerian ring resulting from a uniform
kinematic viscosity is $3\nu$.

\subsection{Governing dimensionless equation}

Once we specify $R$, $P$, $f$ and $\nu$, we have all the ingredients to
solve for the evolution of the ring. To simplify the following
calculations, Eq.~\eqref{SigmaGeneral} is non-dimensionalised. Time is
scaled by $t_\mathrm{e}$ and space by $l_\text{th}$. Surface density is scaled
by the reference density $\sigma_1$. We then define the \emph{dynamical} optical depth via
\begin{equation}
\tau(x,t) = \sigma(x,t)/\sigma_1,
\end{equation}
and so $\sigma_1$ is the density associated with $\tau=1$. 
Note that the dynamical $\tau$ can differ from the ring's physical or photometric optical depth measured by \emph{Cassini}; this is especially the case
when there exist self-gravity wakes (e.g.\ Salo and Karjalainan 2003, Porco et al.~2008, Robbins et al.~2010).
That said, in the regimes relevant to the inner B-ring and the C-ring, the discrepancy is not severe 
and we treat the various optical depths as approximately equal.
Finally, we scale $R$ by $\sigma_1/t_\mathrm{e}$ and $f$ by $1/l_\text{th}$. Both
$R$ and $P$ hereafter will be considered functions of $\tau$.

The scaled evolution equation for $\tau$ is
\begin{equation} \label{taueq}
\d_t \tau = \mathcal{I}-\mathcal{J} +
\tfrac{1}{2}\d_x\left(\mathcal{K}+\mathcal{L}\right)
+ \mu\d_x^2 \tau,
\end{equation}
where we have introduced the ratio of the mass diffusion due to viscosity to
the `ballistic diffusivity'\footnote{If $\nu$ depends on $\sigma$ we can generalise $\mu$, in the linear theory, by
replacing $\nu$ by $d(\nu\sigma)/d\sigma$ evaluated at the unperturbed surface density.}:
\begin{equation} \label{mu}
\mu = \frac{3\nu}{l_\text{th}^2/t_\mathrm{e}}.
\end{equation}
We emphasize that this `ballistic diffusivity' is only a dimensional
estimate, and will see below that ballistic transport does not in fact
lead to a diffusion of mass in the conventional sense.  The four
ballistic transport integrals can be worked into the following forms:
\begin{align}
\mathcal{I}(x,t) &= \int
R[\tau(x-\xi,t)]\,P[\tau(x,t),\tau(x-\xi,t)]\,f(\xi)\,\mathrm{d}\xi,
\label{II}\\
\mathcal{J}(x,t) &= R[\tau(x,t)]\,\int
P[\tau(x+\xi,t),\tau(x,t)]\,f(\xi)\,\mathrm{d}\xi, \label{JJ} \\
\mathcal{K}(x,t) &= \int
R[\tau(x-\xi,t)]\,P[\tau(x,t),\tau(x-\xi,t)]\,\xi f(\xi)\,\mathrm{d}\xi,
\label{KK} \\
\mathcal{L}(x,t) &=  R[\tau(x,t)]\int \, P[\tau(x+\xi,t),\tau(x,t)]\,\xi
f(\xi)\,\mathrm{d}\xi. \label{LL}
\end{align}
If we make the additional approximation that $P$ depends only on the
optical depth at the distant, non-emitting, radius, then we may suppress the
second argument of this function.  This also means that $P$ can be taken
outside the integrals in the expressions for $\mathcal{I}$ and $\mathcal{K}$.
The four expressions can then be written in a compact way, in which $*$ denotes
a convolution integral with respect to $x$:
\begin{align}
&\mathcal{I} = P\cdot (R*f), &\mathcal{J}= R\cdot (P*\tilde f), \\
&\mathcal{K} = P\cdot (R*g), & \mathcal{L} = R\cdot(P*\tilde g).
\end{align}
Here the function $g$ is defined by $g(\xi)=\xi f(\xi)$,
 and the tilde denotes a reflection, so that $f(\xi)=\tilde f(-\xi)$ and
$g(\xi)=\tilde
g(-\xi)$.

\subsubsection{The ratio of mass transport coefficients, $\mu$}

The only control parameter that appears in Eq.~\eqref{taueq} is $\mu$,
which can adopt different values depending on the dominant mode of
viscous transport. In the A-ring, we expect $\nu$ to be monopolised by
the action of self-gravity wakes. Analysis of density wave damping
gives $\nu\approx 30-200$ cm$^2$~s$^{-1}$ (Tiscareno et al.~2007), an estimate
that
 agrees with direct measurements of $\nu$ from $N$-body
simulations of self-gravitating particles (Daisaka et al.~2001, Yasui et al.~2012). In
the inner B-ring and the C-ring, however, we expect little or only moderate
wake activity. In these
cases, estimates
from kinetic theory and simulations give $\nu\sim 0.1 $
cm$^2$~s$^{-1}$ for the C-ring and $\nu \gtrsim 1.0$ cm$^2$~s$^{-1}$ for the
inner
B-ring (Salo et al.~2001; Schmidt et al.~2009, Yasui et al.~2012).

On the other hand, the ballistic diffusivity
$l_\text{th}^2/t_\text{e}$ could be estimated by observing the ranges in
its components from Eqs \eqref{te} and \eqref{lth}.
If $l_\text{th}$ and $t_\text{e}$
are treated as independent, then the ballistic diffusivity is poorly
constrained, its value ranging over four orders of magnitude:
$l_\text{th}^2/t_\text{e}\sim 1-10^4$ cm$^2$~s$^{-1}$.

In fact, $l_\text{th}$
and $t_\text{e}$ are correlated because of the approximate
relation $v_\ee\propto Y^{-1/2}$ (noted in Durisen et al.~1992). Such a
relation would
be consistent with a certain fraction of the impact energy being
transferred to the ejecta. If we denote this fraction by $\epsilon$ we
obtain:
\begin{equation}
\epsilon\, v_\text{imp}^2 = Y\,v_\ee^2,
\end{equation}
where $v_\text{imp}$ is a typical impact speed.
Because $l_\text{th}^2/t_\text{e} \propto Y\,v_\ee^2$, this leads to a
tight bound on the ballistic diffusivity, subject to
an estimate of the transfer efficiency $\epsilon$. Previous numerical
and experimental studies in
a variety of materials, such as basalt, glass, gabbroic anorthosite,
and powdery regolith
(but not ice), yield $\epsilon \sim 0.1$ in the relevant
$v_\text{imp}$ range
(O'Keefe \& Ahrens 1977; Hartmann 1985; Rashev \& Ahrens 2007), and we adopt
this as our
fiducial value.
We then find
\begin{align} 
\frac{l_\text{th}^2}{t_\text{e}} &\approx 1.3\times10^2
\left(\frac{v_\text{imp}}{10\,
    \text{km}/\text{s}}\right)^2\left(\frac{r}{10^5\,\text{km}}
  \right)^3 \notag \\
& \hskip3cm \times \left(\frac{\sigma}{100\,\text{g}/\text{cm}^2}\right)^{-1}\,\text{cm}^2/\text{s}, \label{balldiff}
\end{align}
and we expect it to take values of some $10$ cm$^2$~s$^{-1}$ and $100$ cm$^2$~s$^{-1}$
in the C-ring and in the A and B-rings respectively. Substituting these
estimates into \eqref{mu} yields $\mu\sim 1 $ in
the A-ring and $\mu\sim 0.01$ in
the B and C-rings.

The behaviour of the dimensionless system \eqref{taueq} is thus
controlled by a single tightly constrained parameter, $\mu$. But there still
remains a broad spread in the
physical length and time scales of the problem. Nonetheless
Eqs \eqref{te} and \eqref{balldiff} may replace \eqref{lth}, and
therefore estimations of both the physical length $l_\text{th}$ and
timescales $t_\text{e}$ of
ballistic transport can be reduced to the important
dependency on $Y$, the only poorly constrained parameter. For instance, setting
$Y=10^4$ and
assuming inner B-ring densities
gives $t_\text{e}\sim 10^6$ years and $l_\text{th}\sim 500$~km.

\subsubsection{Probability of absorption}

\begin{figure}
\begin{center}
\scalebox{0.6}{\includegraphics{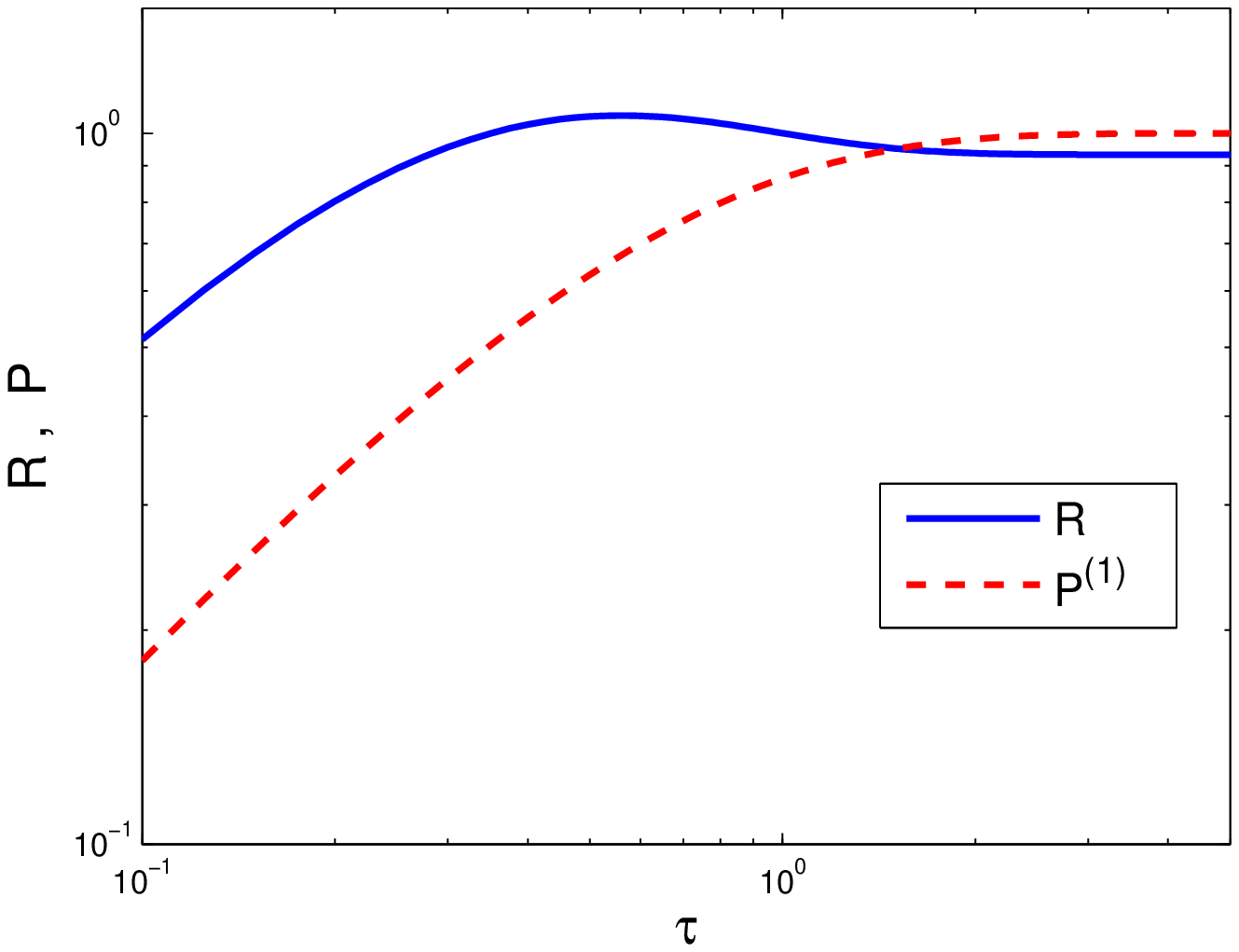}}
\scalebox{0.6}{\includegraphics{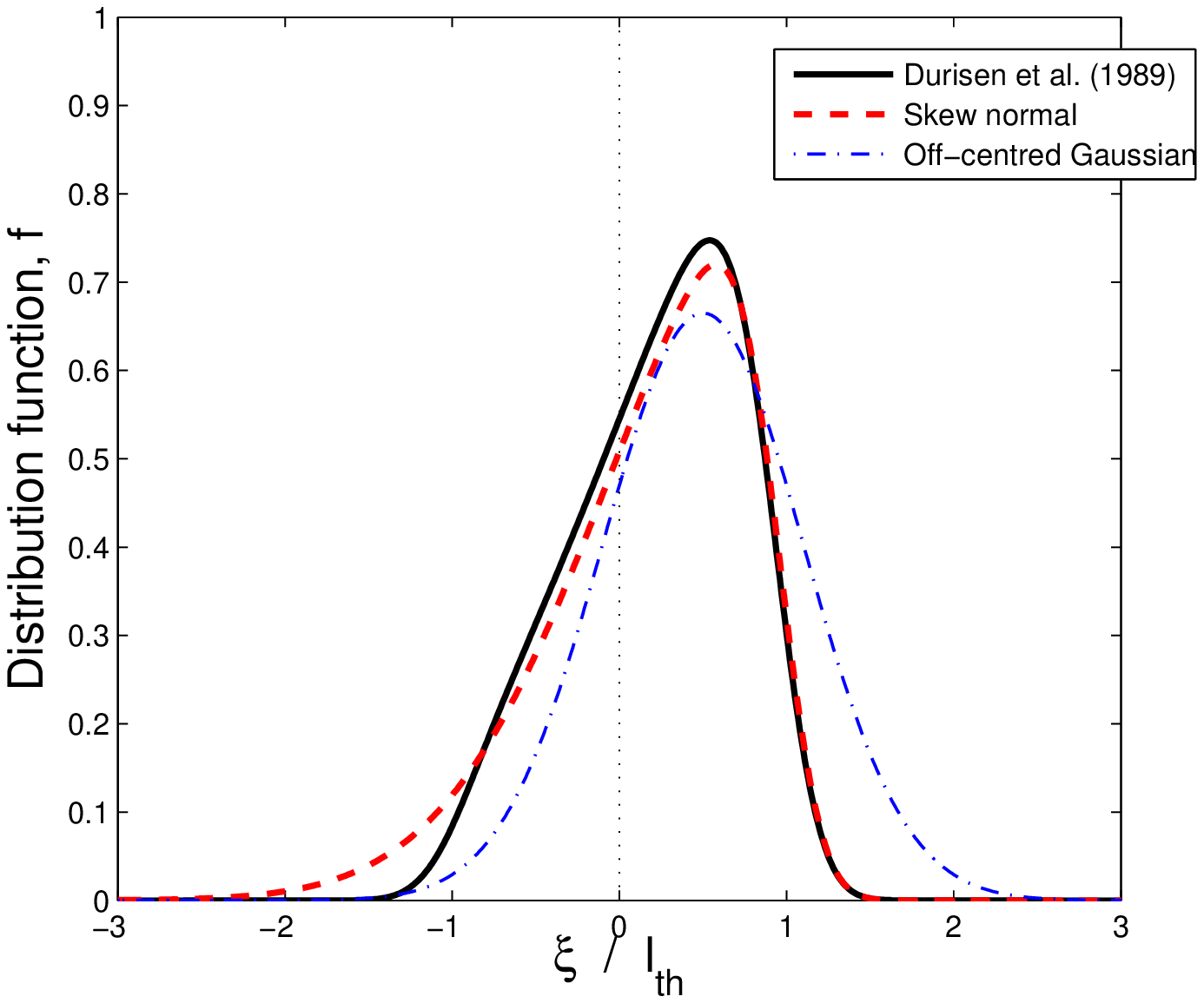}}
 \caption{The first panel shows the normalised mass emission rate $R$ (blue solid line), as approximated by D89 and
   given by Eq.~\eqref{RR}, and the simple absorption probability $P^{(1)}$ (red dashed line), given by Eq.~\eqref{P1}. 
   The second panel shows our approximation
   of the throw
   distribution $f$ as a function of throw distance $\xi$ (black solid
   line) as
   calculated by D89 (see Appendix A).
   In addition we plot more
   convenient analytic estimates: the skew-normal distribution (red dashed
 line) with $a=1$, $\xi_0=1$ and $\beta=-3.5$; and the off-centred Gaussian
 (blue dot-dashed lines)
 with $\xi_0=0.5$ and $a=0.6$.}\label{Rf}
\end{center}
\end{figure}

This and the next subsections present expressions for the functions
$P$, $f$ and $R$ and discuss their form and the relevant physics in
each case.  We begin by describing the probability of absorption $P$. 

The fraction of ejecta that manages to pass through the
ring plane on a given pass is $\text{exp}(-\tau_\text{sl,i})$, where
$\tau_\text{sl,i}$ is the \emph{slant} optical thickness at the radius
of interception. The slant optical thickness will always be equal or
greater than the normal optical depth $\tau_\mathrm{i}$, with the two related
by $\tau_\text{sl,i}= \tau_\mathrm{i}/\cos \theta_\text{inc}$ where
$\theta_\text{inc}$ is the angle of incidence of the ejecta in
  the frame of reference of the receiving ring particles (D89). If we
next suppose that all the ejecta that manage to pass through the ring
at the distant radius are then reabsorbed at the original emitting
radius we may assign
\begin{equation} \label{P1}
P(\tau_\mathrm{i})= 1-\text{exp}(-\tau_\text{sl,i}).
\end{equation}
Where necessary, we denote this expression by $P^{(1)}$ to distinguish
it from the more detailed version presented below.

However, if the emitting region is of low optical depth, an
appreciable amount of material may pass through and undergo another
orbit and hence another opportunity to be accreted at the distant
radius. The probability that ejecta is absorbed on the second try is
\begin{equation}
[1-\text{exp}(-\tau_\text{sl,i})]\exp(-\tau_\text{sl,e}-\tau_\text{sl,i}),
\end{equation}
where $\tau_\text{sl,e}$ is the slant optical depth of the ejecta at the
emitting radius. We now sum over the formally infinite number of
circuits the ejecta may complete and obtain a convergent geometric
series. We can then write down the total probability of absorption:
\begin{equation} \label{P2}
P(\tau_\mathrm{i},\tau_\mathrm{e})=
\frac{1-\text{exp}(-\tau_\text{sl,i})}{1-\text{exp}(-\tau_\text{sl,e}-\tau_\text{sl,i})}.
\end{equation}
Where necessary, we denote this more advanced model by $P^{(2)}$.

To make further progress, we need to account for the angles of
incidence $\theta_\text{inc}$ that appear in formulas \eqref{P1} and
\eqref{P2},
which should vary depending on the geometry of the orbit.
Following D95, we let $\cos \theta_\text{inc}$ take a single `average'
value $\tau_\mathrm{p}$. We thus set $\tau_\text{sl,i} =
\tau_\mathrm{i}/\tau_\mathrm{p}$ and $\tau_\text{sl,e} =
\tau_\mathrm{e}/\tau_\mathrm{p}$. As is argued in D95, $\tau_\mathrm{p}\approx
0.5$ is a
reasonably accurate approximation for a typical set of trajectories, with the error worsening in lower optical depth regions.

\subsubsection{Rate of emission}

In CD90
a detailed formalism is constructed whereby
the intensity and angular distribution of ejecta from a ring layer may
be numerically calculated. The approach borrows much from the
calculation of light-scattering from a layer of particles, treating
the diffuse incident intensity of meteoroids as similar to that of
incident photons. The results of laboratory experiments are also used
to infer the ejecta emission properties, as functions of (a)
the relative velocities of a spherical ring particle and an impacting
 meteoroid, and (b) the angle between the impact direction
and the surface normal.
 The single particle scattering function may
then be
obtained by integrating over the ejecta contributions over all points
on the spherical ring particle. The ejecta distribution function
proceeds directly.
 Finally, though the meteoroid influx is
assumed to be isotropic in the heliocentric reference frame, it is
aberrated in the ring reference frame because of
Saturn's and the ring particles' orbital motion.

The results of these calculations give $R$ as a function of the optical
depth at the emitting radius, and the ejecta distribution $f$ as a
function of the ejection velocity vector. The former may be
approximated, following D95, by the dimensionless analytic form
\begin{equation}\label{RR}
R(\tau) = 0.933\left[ 1 + \left(\frac{\tau}{\tau_\mathrm{s}}-1\right)
\exp(-\tau/\tau_\mathrm{s})\right],
\end{equation}
where $\tau_\mathrm{s}= 0.28$ is a parameter. This expression is normalized
such that $R(1)=1$.

In Fig.~\ref{Rf}a we plot $R$ as a
function of $\tau$. At low $\tau$ the emission rate is small because
the impact rate is small. As $\tau$ increases
so do the number of impacts, and consequently $R$.
 The drop in $R$ at $\tau_0\sim 0.5$ we attribute to a
 transition from an optically thin regime, in which almost all
 ejecta from a given ring particle are thrown into orbit,
 to an optically thick regime, in which an
 increasing amount of liberated ejecta is reabsorbed immediately by
 neighbouring ring particles. 
 At lower $\tau$ it is possible for impact ejecta to leave the ring plane from both sides, whereas at higher $\tau$ 
 ejecta can leave from only one side because of the intervening particles.
  For sufficiently large $\tau$
the emission rate relaxes to a constant value, as incoming
meteoroids penetrate only to an optical depth of order unity
and thus can only dislodge a fixed amount of material.
The tendency for impact ejecta to be immediately accreted by neighbouring particles weakens $R$'s dependence on $\tau$, relative to $P$'s dependence
(see Fig.~\ref{Rf}a). 
This discrepancy plays an important role in the ring's stability and is especially marked near $\tau=0.5$.

\subsubsection{Throw distribution}

The CD90 distribution function is not in the format
required by our formalism as it depends on the ejection velocity:
 the two spherical angles that determine its orientation in the
ring particle frame, $\theta$ and $\phi$, and its magnitude $v_\mathrm{e}$.
The distribution function we
have introduced, instead, depends on a single variable $\xi$, the throw
distance. However, in the context of the transport integrals,
 an approximate mapping between $(\theta,\phi,v_\mathrm{c})$ and $\xi$ is
 possible, which allows us to translate the CD90 distribution to the
 $f$ used in this paper. These details are left to the Appendix.
 In Fig.~\ref{Rf}b, we plot the $f$ constructed in this way alongside two
 convenient analytic approximations, (a) the shifted skew-normal distribution
\begin{equation}\label{skew}
f(\xi)= \frac{1}{\sqrt{2 \pi}}\,
\text{exp}[-(\xi-\xi_0)^2/(2a^2)]\left(1+\text{erf}\left[\beta
\,(\xi-\xi_0)\right]\right),
\end{equation}
where $\text{erf}$ is the error function, $a$ is the `standard
deviation', $\beta$ is the `shape
parameter' which measures the skewness, and the shift is $\xi_0$,
 and (b)
 an off-centred Gaussian,
\begin{equation}\label{Gauss}
f(\xi) = \frac{1}{\sqrt{2\pi a^2}}\text{exp}\left[-(\xi-\xi_0)^2/(2a^2)
\right],
\end{equation}
with shift $\xi_0$ and standard deviation $a$.  Each of these exhibits
the main feature of the realistic distribution, which is its
characteristic asymmetry. Parameters that best match the computed $f$
of D89 are $a=1$, $\xi_0=1$, and $\beta=-3.5$ for the skew normal and $a=0.6$
and $\xi_0=0.5$ for the off-centred Gaussian.

The asymmetry in $f$ plays an important dynamical role,
far greater than the asymmetry arising
from the
 cylindrical
effects that controlled the early studies of edge sharpening and
redistribution
(Ip 1983; Lissauer 1984;
Durisen 1984). This outward bias is a consequence of meteoroids mainly striking the 
leading hemispheres of the ring particles. Ejecta are usually backscattered in these cratering impacts and thus adopt prograde orbits with $v>0$, as in Fig.~\ref{Fig:traject}.
Ring particles suffer more impacts on their leading faces because
of aberration effects that follow from the orbital motion of the ring particles around Saturn and the motion of Saturn itself through the meteoroid flux.
In particular, the orbital motion of the particles gives rise to a `headwind' of matter that increases both the number of impacts on their leading faces and the impact velocities $v_\text{imp}$.
Terrestrial
experiments on ice targets tell us that
the ejecta yield obeys $Y \propto v_\text{imp}^2$ (see CD90), and so the ejecta population (and hence its distribution $f$) 
will be dominated by such prograde ejections.

\subsection{Integral relations}

Before applying our formalism to the question of the ring's stability,
we sketch out some general results regarding the global conservation of mass,
momentum, and energy. Note that the mass and angular momentum of
impacting meteorites have been neglected and thus
do not appear in these balances. This is justified on the basis that
 $Y\gg1$. The injection of energy, however, may be more significant
  and could
 provide a minor source of ring particle velocity dispersion. Our
 formalism, however, does not treat the particles' random motion
 explicitly and this effect appears only through the
 kinematic viscosity $\nu$. See Durisen et
  al.\ (1996) for a more detailed discussion of this physics.

Suppose Eq.~(\ref{dsigmadt}) is solved for a ring of finite radial
extent on an unbounded domain, meaning that either the
density is of compact support, or decays sufficiently rapidly as
$x\to\pm\infty$. The following equations for the moments of the density
distribution can be derived:
\begin{equation}\label{zeroth}
  \frac{\mathrm{d}}{\mathrm{d}t}\int\sigma\,\mathrm{d}x=0,
\end{equation}
\begin{equation}\label{first}
  \frac{\mathrm{d}}{\mathrm{d}t}\int\sigma\,x\,\mathrm{d}x=0,
\end{equation}
\begin{equation}\label{second}
  \frac{\mathrm{d}}{\mathrm{d}t}\int\sigma\,x^2\,\mathrm{d}x=6\nu\int\sigma\,\mathrm{d}x.
\end{equation}
These three equations can be understood as describing the evolution of
the total mass, angular momentum and energy of the ring system.

 It is
clear that Eq.~(\ref{zeroth}) expresses the conservation of mass.
To derive this equation, we note that
\begin{equation}
  \int\mathcal{I}\,\mathrm{d}x-\int\mathcal{J}\,\mathrm{d}x=0,
\end{equation}
which is physically obvious and follows mathematically from a change
of variables in one of the integrals.

As we have already noted, $p_y=\dot y+2\Omega_0x$ plays the role of
specific angular momentum in the local approximation.  For a simple
orbital motion in which $\dot x=0$, $\dot y=-\tfrac{3}{2}\Omega_0x$
and $z=0$, which is the local representation of a circular orbit in
the reference plane, we have $p_y=\tfrac{1}{2}\Omega_0x$.  Therefore
Eq.~(\ref{first}) expresses the conservation of angular momentum.
To derive this equation, we note, using a similar change of variables,
that
\begin{equation}
  \int\mathcal{I}\,x\,\mathrm{d}x-\int\mathcal{J}\,x\,\mathrm{d}x=\int\mathcal{K}\,\mathrm{d}x=\int\mathcal{L}\,\mathrm{d}x,
\end{equation}
and carry out an integration by parts for the viscous term.

The specific energy in the local approximation is
\begin{equation}
\tfrac{1}{2}(\dot
x^2+\dot y^2+\dot
z^2)-\tfrac{3}{2}\Omega_0^2x^2+\tfrac{1}{2}\Omega_0^2z^2,
\end{equation}
which equates to $-\tfrac{3}{8}\Omega_0^2x^2$ for simple orbital motion.
Therefore Eq.~(\ref{second}), which is derived in a similar way,
states that the orbital energy of the ring decreases in time as a
result of viscous dissipation.  Ballistic transport, however,
conserves mass, angular momentum and orbital
 energy.
Eq.~(\ref{second}) also implies that the standard deviation of the
mass distribution of a finite ring increases in time, i.e.\ that the
ring spreads.  This spreading is not (directly) counteracted by
ballistic transport. Thus the edge-sharpening effects observed in simulations should be
understood as a reshaping of a spreading ring feature and may be contrasted to situations where the spreading is actually halted, for instance by a shepherding satellite. 
In such cases Eq.~\eqref{second} is modified by an external torque or angular momentum flux reversal.

The integral relations~(\ref{first}) and~(\ref{second}) do not
generally hold when the equations are solved on a periodic domain because
the angular momentum of material passing through the radial boundaries is not conserved.

\section{Linear stability}

\subsection{Dispersion relation}

Having set up a mathematical formalism
 to tackle the physics of ballistic transport, we now apply it to
 the question of a planetary ring's linear stability. We reserve the
 problems of ring edge sharpening and structure for
future work.

Consider a patch of disk in the homogeneous equilibrium state of
$\tau=\tau_0$ with fixed $\mu$ parameter.
To ease the exposition we assume for the moment
that the absorption probability $P$ depends only on the local optical
depth at the emitting radius, i.e. $P=P^{(1)}$. A disk of uniform $\tau$ is
an equilibrium because $\mathcal{I}=\mathcal{J}$ and the local mass flux
through the domain is uniform. The mass flux is in fact equal to
\begin{equation}
-P_0 R_0 \int \xi f\,\mathrm{d}\xi = -P_0 R_0 \langle \xi \rangle,
\end{equation}
where a subscript 0 indicates evaluation at $\tau_0$, and $\langle \xi
\rangle$ is the expected (mean) throw distance. Note that, in the homogeneous
equilibrium we consider, there is no contribution to the radial drift
velocity from viscous effects. Also note that if the throw
distribution $f$ is completely symmetric there is no drift velocity in
the equilibrium state, as $\langle \xi \rangle=0$. Ballistic transport
then instigates no net angular momentum transport and hence no net
radial drift.

We next superimpose on this state a small perturbation, $\hat{\tau}$. After
linearising Eq.~\eqref{taueq}, we obtain
\begin{align}
&\d_t \hat{\tau} = (R_0 P'_0 - R_0' P_0) \hat{\tau} + R_0' P_0 (\hat{\tau}*f) -
R_0P_0' (\hat{\tau}*\tilde f) \notag\\
&+\tfrac{1}{2}\d_x\left[ (R_0 P_0'+R_0' P_0)
  \langle\xi\rangle\hat{\tau}  + R_0'P_0 (\hat{\tau}*g) + R_0 P_0'
(\hat{\tau}*\tilde g)\right] \notag \\
& \hskip1cm  + \mu\d_x^2\hat{\tau},
\end{align}
where a prime indicates a derivative with respect to $\tau$.
Solutions exist of the form $\hat{\tau} \propto
\ee^{s t+ \iii k x}$, where $s$ is a (complex) growth rate
  and $k$ a (real) wavenumber.  When $\tau$ has this form, the convolutions are
easily evaluated as, e.g.,
\begin{equation}
  \hat\tau*f=F(k)\hat\tau,
\end{equation}
where
\begin{equation}
F(k)= \int f(\xi)\, \ee^{-\iii k \xi}\,\mathrm{d}\xi.
\end{equation}
is the (non-unitary) Fourier transform of $f(\xi)$.  This result can be
seen as a consequence of the convolution theorem, since the Fourier
transform of $\hat\tau$ is proportional to $\delta(k)$.  We thus
obtain the dispersion relation
\begin{align}
&s = R_0 P'_0 - R_0' P_0 + R_0' P_0 F(k) - R_0P_0' F(-k) \notag\\
& +\tfrac{1}{2}\iii k\left[ (R_0 P_0'+R_0' P_0)
  \langle\xi\rangle + R_0'P_0 G(k) + R_0 P_0' G(-k)\right] \notag \\
& \hskip1cm  - \mu k^2,
\end{align}
where $G(k)=\iii F'(k)$ is the Fourier transform of $g(\xi)$, and the
prime indicates differentiation with respect to the wavenumber $k$.
Since $f(\xi)$ is real, $F(-k)=\overline{F(k)}$, where the overline
indicates complex conjugation.  Noting further that $F(0)=1$ and
$G(0)=\langle\xi\rangle$, we find
\begin{equation}\label{dispgen}
s = R_0'P_0\,H(k) - R_0P_0'\,\overline{H(k)} - \mu k^2,
\end{equation}
where
\begin{equation}\label{G}
H(k)= F(k)-1 - \frac{1}{2}k\left[F'(k)+F'(0)\right].
\end{equation}
The first two terms in Eq.~\eqref{G} come from $\mathcal{I}$ and
$\mathcal{J}$ (ballistic mass transport) whereas the bracketed terms
come from $\mathcal{K}$ and $\mathcal{L}$ (ballistic angular momentum
transport). Note that the real part of $H$ comes from the even part of
$f$, while the imaginary part of $H$ comes from the odd part of $f$.
Thus linear modes will manifest as travelling waves unless the
distribution $f$ is completely symmetric, as noted in D95.

\subsection{Stability criterion}

As a consequence of Eq.~\eqref{dispgen}, the real part of $s$ is simply
\begin{equation}\label{Res}
\text{Re}(s) = \left(R_0'P_0 - R_0P_0'
  \right)\text{Re}\left[H(k)\right] - \mu k^2.
\end{equation}
This isolates in a neat mathematical way the various ingredients
governing the mode's potential growth. The first bracketed factor in
the first term accounts for the combined effect of the emission and
absorption rates of the ejecta; the second factor summarises the
influence of the throw distribution; and the last term introduces
viscous damping. It is the first term that is responsible for
instability, and it must arise from either the form of the
absorption/emission profile, the peculiarities of the throw
distribution, or a combination of the two.

`Realistic' distribution functions $f$, such as those given in
Section~2.3.3 and Fig.~\ref{Rf}b, yield $\text{Re}[H(k)]<0$ for \emph{all} $k$.
In
fact, the real part of $H$ can be positive only for exceedingly
asymmetric and/or skewed distributions such as the delta function
(single throw distance)
considered by D95. It follows that instability is controlled by the
first factor in the first term in Eq.~\eqref{Res}. As a consequence, we can
immediately derive
a necessary condition for instability:
\begin{equation} \label{cond}
\frac{d \ln P}{d \tau}> \frac{d \ln R}{d\tau}.
\end{equation}
This condition is satisfied when the
dashed curve in Fig. 3a is steeper than the solid curve (since the
plot is logarithmic).
So in order to obtain a growing mode, the rate of change of absorption must
outstrip
the rate of emission as we increase $\tau$. This makes intuitive sense.
Consider a small overdensity upon a uniform ring: as a result of
the local increase
in $\tau$, both the absorption $P$ and emission $R$ of ejecta will adjust. If
inequality~$\eqref{cond}$ holds then the ring will absorb more ejecta
in relative terms than it releases. As a consequence, material will
start building up at that point and the overdensity will increase,
which in turn will aid the accumulation of even more mass, and so
on. Similarly, in an underdense portion of an otherwise uniform
ring, emission will out-compete absorption and the underdensity will
be exacerbated. The drop in $R$ near $\tau=0.5$, witnessed in Fig.~\ref{Rf}a, 
almost guarantees instability for $\tau\gtrsim 0.5$ because $dP/d\tau>0$.

The stability criterion \eqref{cond} is only a necessary condition
because we have yet to include the damping effect of viscosity. A
sufficient criterion for instability, involving both $\mu$
and $\tau$, is presented in Section 3.3.2. We can, however, establish some
results in the
limit of small and large $k$. At very small scales, i.e.\ large $k$, the
Riemann--Lebesgue
lemma tells us that both $F(k)$ and $F'(k)$ go to zero
for realistic $f$; they do so exponentially fast if $f$ is infinitely
differentiable. Hence the
$-\mu k^2$ term in Eq.~\eqref{Res} dominates in this limit and modes are viscously
damped, as is expected.
On the other hand, on long wavelengths for which $k$ is small,
 a Taylor series expansion reveals that $\mathrm{Re}[H(k)] \propto k^4$.
Therefore
 viscous damping ($\propto -k^2$) will dominate in this limit as well, and
 very long modes will decay. This reflects the fact that ballistic
 transport is weak far beyond its
 throw length. In summary, if the ring is
 unstable, growing modes are limited to a band of intermediate
 wavelengths, as in convection or gravitational instability. In the
 next subsections these results will be illustrated numerically.

In fact the Taylor expansion of $H(k)$ is
\begin{equation}
  H(k)=-\frac{1}{12}F'''(0)k^3-\frac{1}{24}F''''(0)k^4+O(k^5),
\end{equation}
in which $F'''(0)=\iii\langle\xi^3\rangle$ is imaginary and
$F''''(0)=\langle\xi^4\rangle$ is real and positive.  Terms
proportional to $k$ or $k^2$ are absent because of cancellation
between ballistic mass transport and ballistic angular momentum
transport; this result is related to the fact that ballistic transport
conserves the first and second moments of the surface density
distribution (cf.\ Section 2.4).
The absence of a $k^2$ term shows that there is no `ballistic
diffusion' as such,
 although there are dispersion and hyperdiffusion.

Before continuing on to the direct computation of growth rates, we
examine the stability when the
more advanced probability absorption model in Subsection 2.3.1 is
adopted, and $P=P^{(2)}$. Now
$P$ depends on emitting optical depth $\tau_\mathrm{e}$, as well as the
optical depth at the intercepting radius $\tau_\mathrm{i}$:
$P=P^{(2)}(\tau_\mathrm{i},\tau_\mathrm{e})$. The linear analysis can be worked
through as
before, and we find the real part of the growth rate is:
\begin{equation} \label{P2s}
\text{Re}(s) = \left[R_0'P_0 - R_0\,\left(\frac{\d P}{\d
\tau_\mathrm{i}}\right)_0
  +R_0\,\left(\frac{\d P}{\d
\tau_\mathrm{e}}\right)_0\right]\text{Re}\left[H(k)\right] - \mu k^2.
\end{equation}
Here a subscript 0 indicates evaluation at
$\tau_\mathrm{i}=\tau_\mathrm{e}=\tau_0$. 
The
necessary condition for instability, analogous to \eqref{cond}, is
then
\begin{equation}\label{cond2}
 \frac{\d \ln  P}{\d \tau_\mathrm{i}} > \frac{d \ln R}{d \tau} + \frac{\d \ln
P}{\d \tau_\mathrm{e}}.
\end{equation}
Because $P$ is a decreasing function of $\tau_\mathrm{e}$, from Eq.~\eqref{P2},
the additional term on the right-hand side will be negative. As a result,
the instability criterion will be easier to satisfy than \eqref{cond}.
Note that the term in large square brackets in Eq.~\eqref{P2s} is proportional to $-(A-B)$ in D95.

\subsection{Specific examples}

\subsubsection{Growth rates}

We now employ the choices of $P$, $R$ and $f$ introduced in Section~2.3 and
compute growth rates explicitly. The distribution function we
set equal to either the off-centred Gaussian or the skew-normal distribution,
and we mainly
take $P=P^{(1)}$ for simplicity. Unfortunately,
the Fourier transform of the skew normal \eqref{skew} is an
intractable convolution. But the transform of the off-centred Gaussian
\eqref{Gauss}
takes a simple form. In this case
\begin{equation}
F(k)= \text{exp}\left(-\tfrac{1}{2}a^2 k^2 - \ii k \xi_0\right),
\end{equation}
and the real part of $H(k)$ is
\begin{align}
&\text{Re}(H)= -1 + \left[(1+\tfrac{1}{2}a^2k^2)\cos (k\xi_0) \right.\notag \\
&\left.   \hskip2cm+ \tfrac{1}{2}k\xi_0\sin
  (k \xi_0)\right] \text{e}^{-a^2 k^2/2}.
\end{align}
This is negative for all $k \neq 0$ provided that $\xi_0/a \lesssim
4.1595$. In other words, $\text{Re}(H)<0$ unless the distribution $f$
is narrowly confined to a single throw length $\xi_0$.
The best fit to the D89 distribution gives $\xi_0/a \approx
0.83$, which is well within this limit. The single throw
length distribution of $\xi/a \gg 1$ was examined in D95. In this case $H$
is positive and hence the distribution
can drive
instability independently of the $P$ and
$R$ profiles. As a result, the dispersion relation that ensues is more
complicated (but unrealistic).

\begin{figure}
\begin{center}
\scalebox{0.5}{\includegraphics{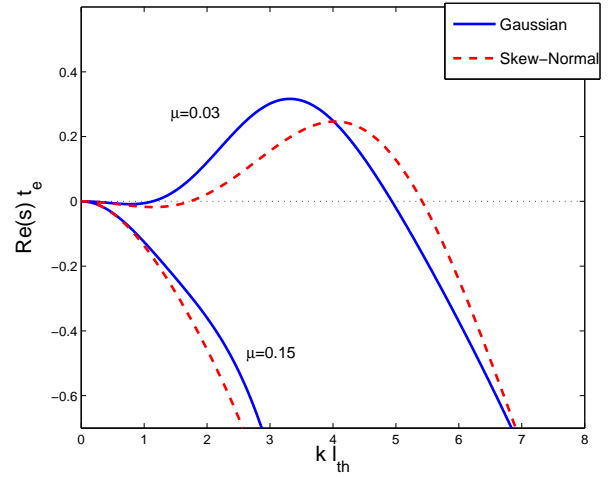}}
 \caption{Real growth rates versus wavenumber $k$ for two values of
   $\mu$ and for two models of the distribution function $f$. The
   solid blue curve represents the off-centred Gaussian of
   \eqref{Gauss} with $\xi_0=0.5$ and $a=0.6$. The dashed red curve
   represents the skew-normal distribution \eqref{skew} with $a=\xi_0=1$ and
   $\beta=-3$. In all cases the background optical depth is
   $\tau=0.5$. The more viscous case $\mu=0.15$ does not exhibit
   instability at any $k$, but for $\mu=0.03$ instability emerges
   on an interval of intermediate wavenumber.}\label{Res1}
\end{center}
\end{figure}

\begin{figure}
\begin{center}
\scalebox{0.6}{\includegraphics{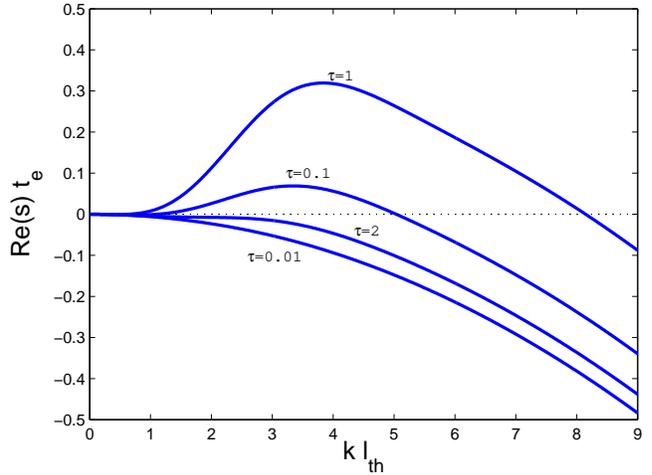}}
 \caption{Real growth rates versus $k$ for various values of $\tau_0$.
 The off-centred Gaussian is used (as in in Fig.~4). The $\mu$
 parameter is held fixed and is equal to 0.006. Note that instability
 occurs only on an intermediate range of optical depth $\tau$; both
 small- and large-$\tau$ rings are stable.}\label{Res2}
\end{center}
\end{figure}

In Fig.~\ref{Res1}, we plot the real part of the growth rate versus the
wavenumber $k$ when
$\tau_0=0.5$ for two values of $\mu$ and for two different distribution
functions $f$. The skew-normal distribution is represented by the
red dashed curve and the off-centred Gaussian by the blue solid
curve. For a lower value of $\mu$, equal to 0.03, both curves
exhibit growth on a characteristic range of wavenumber. Roughly, there is no
growth on scales less than a throw length $l_\text{th}$, nor on scales
longer than about $6l_\text{th}$. The fastest growth occurs on a scale
near $(3/2)l_\text{th}$ with e-folding time
$\sim 4 t_\mathrm{e}$.
As explained earlier, viscous diffusion
dominates both long and short scales, because ballistic transport is
inefficient in each limit. Note, however, that the skew-normal growth curve
exhibits growth on slightly shorter lengths. Otherwise there is good
qualitative agreement between the two distributions. For larger
$\mu$, growth is extinguished as viscous diffusion becomes
 more efficient than ballistic transport and the wave modes
 are smoothed out before they can grow. Note that the growth rates (and their behaviour) are quantitatively consistent with the growth rates computed in D95 (his Fig.~8).

In Fig.~\ref{Res2}, growth rates are given for different $\tau_0$ at a fixed
$\mu$. Here only the off-centred Gaussian has been employed. The
figure illustrates the fact that growth occurs only on intermediate
optical depths, around $\tau_0\lesssim 1$: very large and very
low optical depths do not exhibit instability. This reflects the
relative profiles of the emission and absorption functions $R$ and
$P$ given in Eqs \eqref{P1} and \eqref{RR}. In particular, instability
favours the optical depths where $R$'s rate of change with $\tau$
decreases and becomes negative.

\subsubsection{Stability criterion}

The stability criteria \eqref{cond} and \eqref{cond2} are only
necessary conditions because viscous diffusion has been omitted. A
sufficient condition for instability must incorporate its influence
and thus involve the parameter $\mu$ as well as $\tau_0$. From the
form of the dispersion relation, marginal
stability occurs when Re$(s)= d\text{Re}(s)/dk=0$. Given $\tau_0$,
 we thus have two
equations to solve for a unique critical wavenumber $k$ and critical
$\mu$. This is accomplished numerically, and we plot the ensuing marginal
stability curve $\mu=\mu(\tau_0)$ in the 2D parameter space of $(\mu,\tau_0)$.

\begin{figure}
\begin{center}
\scalebox{0.5}{\includegraphics{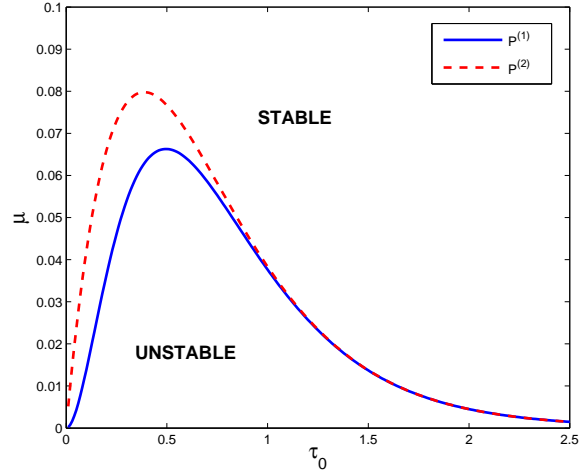}}
 \caption{Curves of marginal stability in the $(\tau_0,\mu)$
   plane. The region below the curve is unstable and the region above
   the curve is stable. The solid blue curve corresponds to the case
   where the simple `one orbit' model $P^{(1)}$ is employed,
   Eq.~\eqref{P1}.
  The dashed curve represents the case where the more
   realistic `multi-orbit' model $P^{(2)}$ is employed,
   Eq.~\eqref{P2}.}\label{stab}
\end{center}
\end{figure}

In Fig.~\ref{stab} we present the marginal stability curves for the two models of
the absorption function $P$, given by \eqref{P1} and
\eqref{P2}. The distribution function $f$ takes the
 off-centred Gaussian profile.
 Regions below the curves
are unstable and regions above the curves are stable.
The red dashed curve represents the more realistic $P^{(2)}$,
 which incorporates the variation in $\tau$ at both
the emission and intersection radii. The blue solid curve
represents the simpler $P^{(1)}$, which accounts only for $\tau$ at the
intersection radius. As is clear, at moderate to large $\tau \gtrsim
1$ the two curves are much the same; this is because most ejecta in
more optically thick regions undergo only one half or one orbit before
being reabsorbed. The $\tau$ at the emitting radius is then
unimportant.
However, at low $\tau$ the two curves deviate, because ejecta
undergo multiple orbits more frequently, an outcome that is not
modelled by the simpler $P^{(1)}$ model. The net effect of these multiple
orbits is to enhance
instability. So for a given low $\tau_0$ the critical $\mu$ can be
double that predicted by the simple model.

Perhaps the most important result here is that the maximum $\mu$
that permits instability is remarkably low ($\approx 0.08$). Viscous
diffusion must be much less than ballistic transport or else
the instability is washed away. Given our estimates for $\mu$,
such a low value immediately rules out the A-ring as a venue for the
ballistic transport instability: we find $\mu$ is $\sim 1$
in the A-ring. On the other hand,
the estimates on $\mu$ for the B and
C-ring ($\sim 0.01$) suggest that instability is possible, but
only barely. From Fig.~\ref{stab}, the critical $\mu$ is 0.032
when $\tau_0=1.1$ and 0.042 when $\tau=0.1$.
If instability occurs in these regions then it may be
near criticality, a fact that should influence its
nonlinear saturation in important ways. We discuss this issue further in Section~4. Finally, we
note that in the outer B-ring, where $\tau_0
> 3$, the critical $\mu$ is tiny, and instability suppressed. 

Generally, for a given (sufficiently small) $\mu$ there exists an interval of
$\tau$ in which instability occurs (cf.\ Fig.~\ref{Res2}, and D95). Instability is suppressed at both high and low
$\tau$. At
high optical thicknesses,
the mechanism of instability becomes weak, as both $P$ and $R$ have similar dependences
on $\tau$ (cf.\
Fig.~\ref{Rf}a). Over and under-densities are only mildly exacerbated, and potentially unstable modes grow too slowly to escape viscous
diffusion. For similar reasons, the instability mechanism weakens at very low $\tau$. 
In this limit as well, both $P$ and $R$ vary similarly with $\tau$ and the first term in \eqref{Res} is small as a result.
Viscous damping again overwhelms
potentially
growing modes. 

These results are consistent with the stability bounds computed in D95. 
If the critical effective yields $Y$ in D95 are translated to critical $\mu$ then the curves in Fig.~7 agree to within a factor of 2 (R.\ Durisen, private communication).
This is encouraging agreement given the different distribution functions used (see the Appendix).

\subsubsection{Phase speeds}

\begin{figure}
\begin{center}
\scalebox{0.6}{\includegraphics{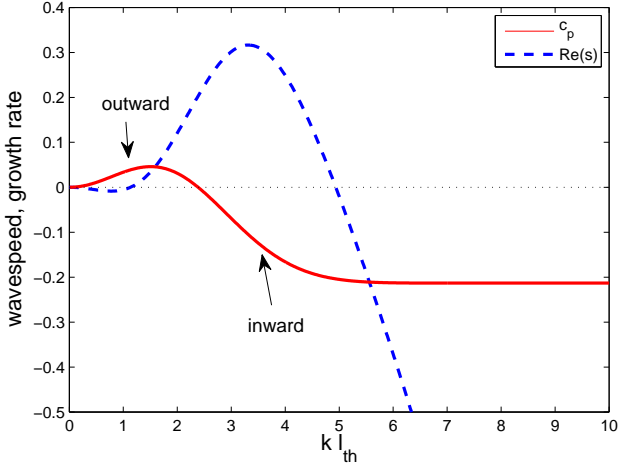}}
 \caption{Phase speed $c_\mathrm{p}$ and real growth rate as functions
   of $k$, for $\tau=0.5$, $\mu=0.01$. The off-centred Gaussian is
   employed for $f$.}\label{cp}
\end{center}
\end{figure}

Generally speaking,
unstable modes manifest as travelling waves, because $s$ is
complex. Only if the distribution $f$ is
perfectly symmetric will modes grow in place. We define a mode's phase
speed $c_\mathrm{p}$ via
\begin{equation}\label{cpp}
c_\mathrm{p} = -\frac{1}{k}\text{Im}(s) = -\frac{1}{k}\left( R\,P
\right)'_0\,\text{Im}[H(k)],
\end{equation}
where we have used the simple model for $P$. In Fig.~\ref{cp} we plot both
the phase speed (red solid curve)
and the real part of the growth rate (blue dashed curve) versus wavenumber
$k$ for the off-centred Gaussian
model of $f$.

A striking feature of $c_\mathrm{p}(k)$ is that it changes
sign as $k$ increases. Longer growing modes propagate outwards, while
shorter growing modes propagate inwards, leaving a critical $k$ at
which a mode grows monotonically. Thus the direction of propagation is not
strictly
tied to the bias in the throw distribution $f$. Having said that, most
modes (including the fastest growing) travel radially inward.
In addition, as $k$
becomes large, $c_\mathrm{p}$ asymptotes to a constant value.

 The sign reversal in the phase speed can be observed mathematically from
 the following argument.  In the limit $k\to0$, we have
 $H(k)\sim-\tfrac{1}{12}\iii\langle\xi^3\rangle k^3$ and so
 $c_\mathrm{p}\sim\tfrac{1}{12}(RP)'_0\langle\xi^3\rangle k^2$.  In the limit
 $k\to\infty$, we have
 $H(k)\sim-\tfrac{1}{2}F'(0)k=\tfrac{1}{2}\iii\langle\xi\rangle k$ and
 so $c_\mathrm{p}\sim-\tfrac{1}{2}(RP)'_0\langle\xi\rangle.$ The relevant throw
 distributions are skewed such that both $\langle\xi\rangle$ and
 $\langle\xi^3\rangle$ are positive, so a sign reversal in $c_\mathrm{p}$ must
 occur at intermediate $k$.

This behaviour can also be understood
in more physical terms. Consider very short modes $k\gg 1$ with
wavelengths much less than both $l_\text{th}$ and the throw length's
standard deviation ($a$ for the off-centred Gaussian).
In such a limit, the influence of the many small undulations are
`averaged away'.
As a result, the mass transport imparts nothing to the collective
motion. However, there will be a net
angular momentum transport which will
excite a net radial flux of material.
 With an outward bias to $f$ this leads to an inward drift.

Consider now a mode with wavelength much longer than
$l_\text{th}$. Ballistic transport is then limited to relatively
short distances. Suppose that mass is thrown almost entirely
 outward and that higher-density regions emit more mass.
A density minimum will increase because
 it is receiving more material from the denser disk inwards to it than
 it can emit.
 Conversely, a density
 maximum will decrease,
 because it is emitting more mass
 than it is receiving from the (less dense) disk inwards to it. 
  As a consequence of this differential mass transport, the
 entire wave-form will crawl outward, in the same direction as
 the throw asymmetry.

The propagation speed of
the fastest growing mode is $\sim 0.2 \, l_\text{th}/t_\mathrm{e}$ which lies
between $0.1$ and $10$ m~yr$^{-1}$. The waves move particularly
slowly. This is consistent with Voyager and Cassini observations of inner
B-ring structure that suggest these undulations have not
travelled appreciably over a 30 year period (Colwell et al.~2009).
According to the above
estimate, a linear mode will have propagated between a few metres and
a few hundreds of metres in that time. The upper limit is just within the
range of detection, but the lower limit certainly is not. The general
consistency is encouraging and suggests a potential constraint on the
typical ballistic transport speed $l_\text{th}/t_\mathrm{e}$. It must be
stressed, however, that the observed structures are most likely
nonlinear waves which may propagate at a different speed from the
linear modes.

\section{Conclusion}

In this paper we construct a mathematical formalism that describes the
ballistic transport in planetary rings. Taking advantage of the
relative smallness of the characteristic throw length $l_\text{th}$,
which is connected directly to the parameter $\varrho \ll 1$, we
employ the local shearing sheet model and manipulate the
transport terms into simple 1D integrals in convolution form. The
resulting main equation is simple to work with, both analytically
and numerically, and its results easy to interpret. Moreover, it is
almost as accurate as the classic formalism of D89, with relative errors
probably of order $\varrho\sim 10^{-3}-10^{-2}$.
In this paper we deploy the model on one facet of the ballistic
transport process: the linear instability (D95). But it may also be
 used to study ramp formation and steep edges.

We derive the linear theory of the instability and apply
our results to Saturn's A, B, and C-rings. The stability analysis can
be framed conveniently in terms of two parameters: $\tau_0$ the background
optical depth  and $\mu$ the ratio of mass transport coefficients due
to viscous diffusion and ballistic
transport. We find that, for
realistic profiles of the absorption probability $P$, the ejecta
emission rate $R$, and the ejecta throw distribution $f$ (D89, CD90),
 instability is pervasive for low and intermediate $\tau$. 
 Actually, instability relies on
 the fact that $R$ increases more weakly with $\tau$ than $P$ does, 
 and it is especially exacerbated by the drop in $R$ for
 $\tau \gtrsim 0.5$ (cf.\ Fig.~\ref{Rf}a). Near this optical depth, a small increase in density
 instigates a fall in emission, and a concurrent rise in
 absorption. The small overdensity is hence reinforced and the process
 runs away. The drop in $R$ at this $\tau_0$ we attribute to a
 transition from an optically thin regime in which almost all
 ejecta from a given ring particle are thrown into orbit,
 to an optically thick regime in which
 increasing amount of liberated ejecta is reabsorbed immediately by
 neighbouring ring particles.

However, the ballistic transport instability is vulnerable to viscous diffusion
which can smear out potentially growing modes. In fact, the critical
$\mu$ above which instability is always extinguished is low:
$\mu_\mathrm{c}\approx 0.08$. In the A-ring,
self-gravity wakes dominate and viscous diffusion is relatively efficient; here $\mu$ takes values $\sim 1$ and instability never occurs. 
In the B-ring, $\mu\sim 0.01$, which precludes instability in its extremely dense outer regions, where $\tau>3$. 
For these large optical depths, Fig.~\ref{stab} indicates $\mu_\mathrm{c}$ is tiny\footnote{
The large-scale 100-km structure observed in the outer ring is most likely generated by something other than ballistic transport, 
perhaps electromagnetic instability (Goertz and Morfill 1988, Shan and Goertz 1991).}. On the other hand, the inner regions of the B-ring exhibit
$\tau_0\approx 1$ with a corresponding critical value of
$\mu_\mathrm{c}\approx 0.032$. Thus instability can occur but should be near
marginality.
The case is the same in the C-ring, where at $\tau_0=0.1$ we have
$\mu_\mathrm{c}\approx 0.042$.

 While we can tightly constrain the governing
stability parameters, $\mu$ and mean optical depth $\tau_0$, the
physical length and time scales of the phenomena are less easy to tie
down, owing to uncertainties in the physical state of ring particle
surfaces (and hence $Y$). Thus
an exact determination of the preferred length-scales of the structures cannot
yet be made.
However, we may be able to better constrain these
by matching the results of nonlinear simulations with
the observed structures in detail.

The fact that instability may be near marginality in both the B and
C-rings may have important dynamical consequences. On the one hand,
instability may saturate at a low amplitude. Indeed the C-ring
exhibits gentle undulations of
1000-km wavelength. The C-ring, however, also supports 100-km plateau
structures of
relatively large amplitudes at slightly larger radii (Colwell et
al.~2009).
Can the
ballistic transport instability generate both sets of structures
concurrently? This might suggest that ring properties in the C-ring
vary rapidly with radius, yielding $l_\text{th}\sim 1000$~km near
$r=80000$~km and $l_\text{th}\approx 100$~km at $r=90000$~km. But it is
also possible that the nonlinear dynamics permits the coexistence of
both dominant lengthscales; if so, detailed numerical simulations may help in
showing how.

On the other hand, a system near marginality could exhibit
bistability, whereby both a linearly stable homogeneous state and an active
large-amplitude state are supported. In fact, the inner B-ring
displays adjoining flat zones and active `wave zones' between
$r=93000$~km and $r=98000$~km (Colwell et al.~2009). Can we associate
these two regions with the inactive (homogeneous) and active (wavelike) states of a bistable
system? If so, what determines the arrangement of the states; do the
states propagate into each other at a `front'; and at what speed would
this occur? Alternatively, the ring viscosity may vary sufficiently over the inner B-ring to permit instability
in some regions and not in others. The main questions are then: what is the cause of the large-scale viscosity variation and how much would it need to change to explain the observations?
Because the system is near marginality here, the viscosity need not change by a great amount. 

Some of these questions we hope to answer in future work where we
will present the nonlinear theory and simulations of the instability. There
we will exhibit the formation of wavetrain solutions, their stability to
secondary modes, and the dynamics of potentially bistable regions of
the disk. These various behaviours and patterns we will subsequently
connect to observations.

\section*{Acknowledgments}
The authors would like to thank the reviewer Paul Estrada for a set of helpful comments that
much improved the paper. The final manuscript also greatly benefited from discussions with Richard Durisen and his 
extensive and insightful comments on an earlier version of this work.
This research was supported by STFC grants ST/G002584/1 and ST/J001570/1.

\appendix

\section{Connection with the D89 formalism}

Most of the theoretical apparatus of D89 can be massaged into the form we
present here by taking the local approximation, expanding in the small
parameter $\varrho$ and retaining only leading-order terms. A number of
additional
assumptions then need to be made in order to connect the loss and gain
integrals of D89 with the simpler versions that appear in Section~2. We sketch
out these details now.

In the formalism of D89 the ballistic transport terms are generally of
the form
\begin{equation}
\Lambda= \int_0^\infty \int_0^{2\pi} \int_{-1}^{1}
\Theta(\tau_\text{sl,i})\,g(\varrho,\phi,\theta)\,d\varrho\,d\phi\,d\cos\theta,
\end{equation}
where the angles $\phi$ and $\theta$ delineate the ejecta velocity
orientation and $\varrho$ its magnitude (in units of $r\Omega_0$). The
angle $\theta$ denotes the angle between the $y$ axis and the ejecta
velocity, and $\phi$
the angle between $z$ axis and the projection of the
velocity on to the $(x,\,z)$ plane (see Fig.~2 in D89). The
function $\Theta$ is either $P$ or $R$, and $g$ is the CD90
ejection distribution function. Recall that $\tau_\text{sl,i}$ is the slant
optical
depth of the ring at the distant radius of ejecta re-intersection.
The potential dependence of $P$ on $\tau_\text{sl,e}$ is
not denoted for ease of presentation.

First we make the approximation
$\tau_\text{sl,i}=\tau_\mathrm{i}/\tau_\mathrm{p}$, that is we
treat the slant optical depth as a simple function of normal optical
depth only, as assumed in
Section~2.3.1. We then can set $\tau_\mathrm{i}= \tau(r_\mathrm{i})$, where
$r_\mathrm{i}$ is the
radius of the distant reintersection point. This radius may be
expressed in terms of the ejecting radius $r_\mathrm{e}$ and
$(\varrho,\phi,\theta)$:
\begin{equation}
r_\mathrm{i}= \left[\frac{1+ 2 \varrho \cos\theta +
  \varrho^2(\cos^2\theta+\sin^2\theta\cos^2\phi)}
{1-2\varrho\cos\theta
-\varrho^2(\cos^2\theta+\sin^2\theta\cos^2\phi)}\right]r_\mathrm{e},
\end{equation}
(see D89 and Durisen et al.~1996). Expanding in small $\varrho$ gives simply
\begin{equation}
r_\mathrm{i}=r_\mathrm{e}\left(1+4\varrho\cos\theta + \mathcal{O}(\varrho^2)
\right).
\end{equation}
Thus if we retain only terms up to $\varrho$ then
$r_\mathrm{i}=r_\mathrm{i}(\varrho,\theta)$ and hence
$\Theta=\Theta(\varrho,\theta)$ and
the $\phi$ dependence vanishes. All that matters now is the component
of the ejecta velocity in the $y$-direction.

Next we define the throw distance
$$\xi= r_\mathrm{i}-r_\mathrm{e} = 4\varrho r_\mathrm{e}\,\cos\theta,$$
transform the $\theta$
 integral into a $\xi$ integral, and swap the order of
 integration. This yields the form familiar from Section~2,
\begin{equation}
\Lambda= \int_{-\infty}^\infty \Theta[\tau_\mathrm{i}(r_\mathrm{i}+\xi)]\,
f(\xi)\, \mathrm{d}\xi,
\end{equation}
where we have introduced the new distribution function
\begin{equation} \label{Durif}
f(\xi)= \int_0^{2\pi} \int^\infty_{|\xi|/(4 r_\mathrm{e})} \frac{g}{4\varrho
r_\mathrm{e}}\, d\phi\,d\varrho.
\end{equation}

D89 introduce an analytic approximation to the full numerical solution
of $g$. It takes the form
\begin{equation} \label{ggg}
g = |\sin\phi|^{1/4}\,\theta\,\text{exp}(-\theta^{1.3})\,h(\varrho),
\end{equation}
where $h$ is a yet to be specified function of $\varrho$. Using this
estimate, the $\phi$ integral in \eqref{Durif} can be done
immediately, leaving the sole $\varrho$ integral. The simplest choice
for $h$ is a delta function $\delta(\varrho-\varrho_\mathrm{c})$, for fixed
$\varrho_\mathrm{c}$. This means that all ejecta are released with the same
speed. This then accounts for the $\varrho$ integral and we obtain
\begin{equation}
f \propto
\cos^{-1}(\xi/l_\text{th})\,\text{exp}[-(\cos^{-1}(\xi/l_\text{th}))^{1.3}],
\end{equation}
where we have set $l_\text{th}= 4 \varrho_\mathrm{c} r_\mathrm{e}$.
A more realistic choice but with the same basic properties is
\begin{equation}
h(\varrho) = \varrho
\,\text{exp}[-(\varrho-\varrho_\mathrm{c})^2/(2a_\mathrm{D}^2)],
\end{equation}
 for some (narrow) dispersion $a_\mathrm{D}^2$. The $\varrho$ integral in
 \eqref{Durif} then must be accomplished numerically. The resulting
 function is plotted in Fig.~\ref{Rf}b with $a_\mathrm{D}=0.2$.
Note that in D95, 
$g$ is mainly taken to be a distribution uniform over hemispheres with $h$ a truncated power law, derived from hypervelocity experimental data (Durisen et al.~1992).

\end{document}